\documentclass[twocolumn,pra,preprintnumbers,amsmath,amssymb,superscriptaddress]{revtex4}
\newcommand{\p}{\partial}

\begin{document}

\preprint{CERN-PH-TH-2009-205}
\preprint{RIKEN-TH-175}

\title{$N$-Body Nuclear Forces at Short Distances in Holographic QCD}

\author{Koji {\sc Hashimoto}}\email[]{koji(at)riken.jp}
\affiliation{
{\it Nishina Center, RIKEN, Saitama 351-0198,
Japan }}

\author{Norihiro {\sc Iizuka}}\email[]{norihiro.iizuka(at)cern.ch}
\affiliation{
{\it 
Theory Division, CERN, CH-1211 Geneva 23, Switzerland}}

\author{Takashi {\sc Nakatsukasa}}\email[]{nakatsukasa(at)riken.jp}
\affiliation{
{\it Nishina Center, RIKEN, Saitama 351-0198,
Japan }}

\begin{abstract}
 We provide a calculation of $N$-body ($N\ge 3$) nucleon interactions at
 short distances in holographic QCD. In the Sakai-Sugimoto model of
 large $N_c$ massless QCD, $N$ baryons are described by $N$ Yang-Mills
 instantons in 5 spacetime dimensions.  
 We compute a {\it classical} short distance 
 interaction hamiltonian for $N$ 'tHooft instantons. This corresponds to
 $N$ baryons  sharing identical {\it classical} spins and isospins. 
 We find that genuine $N$-body nuclear forces turn out
 to vanish for $N\ge 3$, at the leading order. This suggests
 that classical $N$-body forces are always suppressed compared with
 2-body forces.  
\end{abstract}

\maketitle

\noindent
--- {\bf Introduction.} 
Recent developments in computational nuclear physics reveals that
    the microscopic description of nucleus in terms of nucleon
    degrees of freedom requires three-nucleon interactions.
In fact, 
although the nuclear three-body interaction is weaker than
the two-body interaction, the binding energies of light nuclei \cite{1} 
and
the saturation density of nuclear matter \cite{2} 
cannot be understood without
taking into account the three-body terms. This is due to a large
cancellation of the kinetic energy and two-body attraction. The main
component of the three-body interaction is associated with two-pion
exchange, such as the Fujita-Miyazawa force \cite{3}. 
However, in addition to
this, a repulsive three-body interaction of short range is required for
quantitative description of nuclear systems \cite{1,4}. The short-range
three-nucleon interaction, which is assumed to be spin-isospin independent
in many cases, is important for determination of the nuclear equation of
state at high density \cite{2,5}.

The two-body interactions adopted in those many-body calculations
    are determined by the phase-shift analysis of nucleon-nucleon
    scattering data. However, much less information is available
    for the $N$-body forces ($N\geq 3$). Of course, we know that, in
    principle, the nuclear properties should be derived from QCD
\cite{Ishii}.
    However, QCD is strongly coupled at the nuclear energy scale, which
    leads to a huge gap between QCD and nuclear many-body problems.

A recent progress in string theory can bridge this gap, analytically. 
It is called holographic QCD, 
an application of of gauge/string duality \cite{Maldacena:1997re} to 
strongly coupled QCD. We apply the holographic QCD 
to $N$-body nuclear force ($N\geq 3$). 

In holographic QCD, 
one of the most successful D-brane models 
is 
Sakai-Sugimoto model (SS model)\cite{Sakai:2004cn,Sakai:2005yt}. 
The theory, which is a $U(N_f)$
Yang-Mills-Chern-Simons (YM-CS)
theory in a warped 5-dimensional space-time,
was conjectured to be dual to low energy massless QCD with $N_f$
flavors, in the large $N_c$ and large $\lambda$ limits 
($\lambda \equiv N_c g_{\rm QCD}^2$ 
is a 'tHooft coupling of QCD).
Modes of the gauge fields correspond to meson degrees of freedom and  
this model reproduces surprisingly well various expected features of 
hadrons, incorporating  very nicely the nature of chiral Lagrangians. 

Baryons are identified with soliton solutions localized in the spatial
4-dimensions \cite{Sakai:2004cn}. This is  
quite analogous to that in pion effective theory, 
baryons are identified with Skyrmions.
Quantization of a single soliton in the SS model
\cite{Hata:2007mb,Hata:2007tn} 
gives baryon spectra, and also chiral properties 
such as charge radii and magnetic moments 
\cite{Hashimoto:2008zw} (for other approaches to baryons, see 
\cite{Hong:2007kx}). 
Meson-baryon-baryon couplings \cite{Hashimoto:2008zw} give a basis of
a 2-body nuclear force at long distances, {\it a la} one-meson-exchange
picture. Short-distance nucleon-nucleon forces were computed 
\cite{Hashimoto:2009ys}, 
which generates a repulsive core    
with analytic formula for potentials in the large $N_c$ limit.
A key is that the warping can be absorbed into the
rescaling of the YM-CS theory and brings the string scale to QCD scale. 
Furthermore, 
when two solitons are close to each other, the warping factor is almost
constant, therefore the effects
of the curved geometry can be ignored 
so that an exact two-soliton solution is available.  

In this letter, 
we compute $N$-body nuclear forces for arbitrary $N$,
with {\it exact} $N$-instanton solutions, 
generalizing the method in Ref.~\cite{Hashimoto:2009ys}.
The exact treatment is in contrast to
the Skyrmion and other chiral soliton models,  
in which 
multi-soliton solutions are quite difficult to obtain.


\vspace{2mm}
\noindent
--- {\bf Nuclear Force at Short Range.} 
Baryons, including nucleons, are identified with solitonic solutions in
the SS model \cite{Sakai:2004cn}, and 
we provide a brief review of the construction of the soliton and 
the 2-body nuclear force at short range for two flavors ($N_f=2$)
computed in \cite{Hashimoto:2009ys}. 

The following rescaling of the coordinates \cite{Hata:2007mb} 
can allow one to understand the system as a $1/\lambda$
perturbation around 
a flat space, which is suitable for studying the instanton
solution: 
\begin{eqnarray}
 \widetilde{x}^M = \lambda^{1/2}x^M \quad (M=1,2,3,4), \quad
\widetilde{x}^0 = x^0,
\end{eqnarray}
and accordingly 
$\widetilde{A}_0(t,\widetilde{x}) = A_0(t,\widetilde{x})$ and 
$\widetilde{A}_M(t,\widetilde{x})=\lambda^{-1/2}A_M(t,\widetilde{x})$. 
In the following we omit the tilde for simplicity.
In these new variables, there are essentially two
deviations from the YM theory in the flat space, at the leading order
in $1/\lambda$ expansion: 
(i) the effect of the
CS term, and (ii) the effect of the space weakly curved along the $x^4$ 
direction. The additional hamiltonians are
\begin{eqnarray}
 H_{\rm pot}^{U(1)} &\equiv&
\frac{-a N_c}{2}
\int\! \prod_M dx^M \;  \hat{A}_0 \Box \hat{A}_0 ,
\label{a0a0}
\\
H_{\rm pot}^{SU(2)}&\equiv& 
\frac{a N_c}{6}\int\! \prod_M dx^M \; (x^4)^2 \; {\rm tr} (F_{MN})^2,
\label{su2h}
\end{eqnarray}
respectively, with $a \equiv 1/(216 \pi^3)$. 
Note that we work in the unit $M_{\rm KK}=1$, where 
$M_{\rm KK}$ is the unique scale parameter appearing in the model,
and it can be fixed by fitting the $\rho$ meson mass, giving
$M_{\rm KK}=949$ [MeV].

In particular, for a single baryon, the leading order solution 
is a single instanton in the flat 4-dimensional space, 
that is BPST instanton \cite{Belavin:1975fg}.
The instanton has moduli parameters: the instanton location
$X^M$, the size $\rho$ and the orientation in $SU(2)$.
These hamiltonians induce potentials in the 
moduli space of the instanton, and 
$\rho$ and $X^4$ prefer particular values classically,
$(\rho_{\rm cl})^2 =\frac{1}{8\pi^2 a}\sqrt{\frac{6}{5}}$, 
$X^4_{\rm cl}=0$.
For multi-instantons, 
there appears a potential for the moduli representing the 
distance between the
instantons, which is in fact the nuclear force. 

In Ref.~\cite{Hashimoto:2009ys}, this 2-body 
nuclear force was evaluated explicitly.
Multi-instanton solutions are available in flat space, while in this
particular curved space it is difficult to find them. However, when
instantons are close enough to each other, the effect of the curved
space can be neglected, and as a leading order solution we can use the
multi-instanton solutions in the flat space. Therefore, the distance
$r_{ij}$ between the $i$-th and the $j$-th nucleons 
allowed in this approximation is 
$|r_{ij}| <  M_{\rm KK}^{-1}$ 
($|r_{ij}| < \lambda^{1/2}M_{\rm KK}^{-1}$)
in the original (rescaled) coordinates. 
Thus we probe only the short range for the nuclear force.


The construction of the two-instanton solution owes to the renowned ADHM
(Atiyah-Drinfeld-Hitchin-Manin) 
method \cite{Atiyah:1978ri, Corrigan:1983sv}. 
The moduli parameters of generic $N$ instanton solutions are completely
encoded in the real $N\times N$ matrix function $L(x ; X,\cdots)$.
The Osborn's formula \cite{Osborn:1979bx}
tells us the instanton density 
\begin{eqnarray}
 {\rm tr} (F_{MN})^2 = \Box^2 \log\det L
\end{eqnarray}
where 
$\Box \equiv \partial_M \partial_M$.
Using this expression, the equation of motion for the 
U(1) part of the gauge field 
which is sourced by the instanton density is solved as
\cite{Hata:2007mb} 
\begin{eqnarray}
 \hat{A}_0= \frac{1}{32\pi^2 a} \Box \log \det L \, .
\end{eqnarray}

With this explicit dependence on the instanton moduli parameters
in $L$, one can compute the
hamiltonians (\ref{a0a0}) and (\ref{su2h}) as functions of them.
Then, the expectation value of the hamiltonians for given baryon states
(the wave functions are written by the moduli parameters)
gives the nuclear force at short range \cite{Hashimoto:2009ys}.

There is the third contribution to the additional hamiltonians,
$H_{\rm kin}$, 
which is present only in multi-instanton case. This comes
from the metric of the instanton
moduli space. In Ref.~\cite{Hashimoto:2009ys}, it was shown that
it is higher order in $1/N_c$
compared to the other two hamiltonians (\ref{a0a0}) and (\ref{su2h}), 
so we need not compute it in this paper. 


\vspace{2mm}
\noindent
--- {\bf 3-body Nuclear Force.} 
The 2-body nuclear force computed in Ref.~\cite{Hashimoto:2009ys} is 
for generic spin/isospin components. 
But since explicit generic $N$ instanton solution is not available, we
consider a special solution called 'tHooft instanton which has $5N-3$ 
moduli parameters (while generic instanton solution has $8N-3$ moduli
parameters). It is important to notice that once we restrict our moduli
space by hand like this, we cannot get the generic expression for the
nuclear force for given baryon states. Instead, what we will
obtain is a classical analogue of the nuclear force. 

The moduli parameters of the 'tHooft instantons are 
only the size $\rho_i$ and the location 
$X_i^M$ of each instanton ($i=1,2,\cdots,N$).
The missing parameters, the
orientations of the instantons in $SU(2)$,
are responsible for the spin/isospin 
wave functions of the baryons. 
Thus our analysis with the 'tHooft instantons is restricted to
``classical'' baryons, where all the spin/isospins of the
baryons are identical classically.

First, let us show that 
$H_{\rm pot}^{SU(2)}$ given in Eq.~(\ref{su2h}) is irrelevant to the
three-body nuclear
forces.
We can use the generic formula obtained 
in Appendix $C$ of Ref.~\cite{Hashimoto:2009ys},
\begin{eqnarray}
 \int d^4x \; (x^4)^2 \; {\rm tr} (F_{MN})^2 = 8 \pi^2
\sum_{i=1}^N\left(2(X^4_i)^2 + \rho_i^2\right),
\end{eqnarray} 
for the $N$ 'tHooft instantons.
The expression consists of just a sum of each instanton sector,
which means that there is no term involving the inter-nucleon distance,
that is, no contribution to the nuclear force.
Therefore, we compute the other hamiltonian
(\ref{a0a0}) in this paper. (The contribution from $H_{\rm kin}$
is suppressed as in the case of the 2 instantons \footnote{
Let us explain briefly why $H_{\rm kin}\equiv-\nabla^2/16 \pi^2 a N_c$
gives smaller contributions compared to the other two hamiltonians.
Here $\nabla^2$ is 
the metric on the multi-instanton moduli space.
We  evaluate the expectation value of $\nabla^2$
with the baryon states. The dimension of $\nabla^2$
is [length]$^{-2}$. The leading 2-body forces
are $\propto 1/X_{12}^2$,
so the remaining factor should be dimensionless.
Thus $\nabla^2$ originating in the metric is written only by using
a dimensionless operator
$y\cdot \partial/\partial y$ twice, where $y$ is the moduli $\rho$ and
the SU(2) orientations.
Then it was found in Ref.~\cite{Hashimoto:2009ys} that 
$\langle \nabla^2\rangle={\cal O}(N_c)$,
so, in total, after rescaling back $|X|^2\to \lambda |X|^2$, one obtains
$\langle H_{\rm kin}\rangle = {\cal O}(\lambda^{-1})$,
which is smaller than 
$\langle H^{SU(2)}\rangle \sim 
\langle H^{U(1)}\rangle \sim {\cal O}(N_c/\lambda)$
by the factor of $1/N_c$.
This suppression by $1/N_c$ is expected to any instanton number,
that is, general $N$-body force, and so in this paper we don't consider
$H_{\rm kin}$.}.)

In this section, we concentrate on 
the case for $N=3$, {\it i.e.}~the 3-body force. 

For three 'tHooft instantons, which correspond to nucleons sharing 
classically 
identical spins/isospins, we
have 
\begin{eqnarray}
L 
= \left(
\begin{array}{ccc}
(x\!-\!X_1)^2\!+\!\rho_1^2 & \rho_1\rho_2 & \rho_1 \rho_3 \\
\rho_1 \rho_2 & \!\!\!(x\!-\!X_2)^2\! +\! \rho_2^2 & \rho_2 \rho_3\\
\rho_1 \rho_3 & \rho_2 \rho_3 & \!\!\!(x\!-\!X_3)^2\! +\! \rho_3^2
 \\
\end{array}
\right),
\end{eqnarray}
where we omit the index $M$ and denote $(x^M)^2$ by $x^2$.
Then 
the Osborn's formula becomes particularly simple, 
\begin{eqnarray}
 \log \det L = \sum_i \log (x-X_i)^2 + \log f \,, 
\label{logdetl}
\end{eqnarray}
with $f \equiv 1 + \sum_i\frac{\rho_i^2}{(x-X_i)^2}$.
This gives  the $U(1)$ gauge field 
\begin{eqnarray}
\hat{A}_0 = 
\frac{1}{32\pi^2 a} 
\left[\Box \sum_i \log (x-X_i)^2 + 
\frac{\Box f}{f} - \frac{(\p_M f)^2}{f^2}
\right] \, .
\label{last}
\end{eqnarray}
The first term in Eq.~(\ref{logdetl}) is a self-energy 
which was already computed, and $f$ is a harmonic function,
$i.e.$, $\Box f=0$. Thus, all we need to evaluate is
only
the last term in Eq.~(\ref{last}), $ (\p_M f)^2/f^2$.

For three instantons, we can expand the expression for 
$(x-X_1)^2 \ll (x-X_2)^2, (x-X_3)^2$.
In particular, we can approximate
$(x-X_2)^2\sim (X_1-X_2)^2 \equiv X_{12}^2$,
and a similar expression for $X_{13}$. 
Furthermore, for simplicity we put $X_1^M=0$.
Then, the expansion is
\begin{eqnarray}
&&\hat{A}_0 = 
\frac{1}{8\pi^2 a} 
\left[
\frac{1}{x^2}\left(1-\frac{\rho_1^4}{(x^2 + \rho_1^2)^2}\right)
+ \frac{1}{X_{12}^2}+ \frac{1}{X_{13}^2}
\right. \nonumber \\
&&+
\frac{2\rho_1^4}{(x^2\! +\! \rho_1^2)^3}
\left(\!\frac{\rho_2^2}{X_{12}^2}\! +\! \frac{\rho_3^2}{X_{13}^2}\!\right)
\!+\! 
\frac{-3\rho_1^4 x^2}{(x^2\! +\! \rho_1^2)^4}
\left(\!\frac{\rho_2^2}{X_{12}^2}\! +\! \frac{\rho_3^2}{X_{13}^2}\!\right)^2 
\nonumber \\
&&\left.
+
\frac{-2\rho_1^2}{(x^2\! +\! \rho_1^2)^2}
\left(\frac{\rho_2^2 x\cdot X_{12}}{X_{12}^4} 
\!+\! \frac{\rho_3^2 x\cdot X_{13}}{X_{13}^4}\right)
\!+\! {\rm higher}
\right] \,.
\end{eqnarray}

We like to compute the potential (\ref{a0a0}).
As seen from the expression for $\hat{A}_0$, 
the leading term of the nontrivial three-body force has the form 
$\frac{1}{X_{12}^2 X_{13}^2}$.
This means that, 
we should expect
the nuclear force appearing {\it in proper to the 3-body} 
is, at the leading order,
\begin{eqnarray}
 H = {\cal O}\!\left(\frac{N_c}{\lambda^2}\right)
 \frac{1}{X_{12}^2 X_{13}^2} \, ,
\label{leadform}
\end{eqnarray}
in the original coordinates.
Let us consider only terms of this leading form (\ref{leadform}). 
We obtain
\begin{eqnarray}
&&H_{\rm pot}^{U(1)} \biggm|_{\rm 3-body} = 
\frac{-a N_c}{2 (8\pi^2a)^2}  \nonumber \\
&&\times  \int\! d^4x \;  
\left[
\frac{1}{x^2}\left(1-\frac{\rho_1^4}{(x^2 + \rho_1^2)^2}\right)
\Box
\frac{-6\rho_1^4 x^2}{(x^2 + \rho_1^2)^4}
\frac{\rho_2^2\rho_3^2}{X_{12}^2X_{13}^2}
\right.
\nonumber 
\\
&&
+
\frac{-6\rho_1^4 x^2}{(x^2 + \rho_1^2)^4}
\frac{\rho_2^2\rho_3^2}{X_{12}^2X_{13}^2}
\Box
\frac{1}{x^2}\left(1-\frac{\rho_1^4}{(x^2 + \rho_1^2)^2}\right)
\nonumber 
\\
&&
+
\frac{1}{X_{12}^2 X_{13}^2}
\left(1 + \frac{2\rho_1^4 \rho_2^2}{(x^2 + \rho_1^2)^3}\right)
\Box
\left(1 + \frac{2\rho_1^4 \rho_3^2}{(x^2 + \rho_1^2)^3}\right)
\nonumber 
\\
&&
\left.
+
\frac{1}{X_{12}^2 X_{13}^2}
\left(1 + \frac{2\rho_1^4 \rho_3^2}{(x^2 + \rho_1^2)^3}\right)
\Box
\left(1 + \frac{2\rho_1^4 \rho_2^2}{(x^2 + \rho_1^2)^3}\right)\right]
\nonumber 
\\
&&+ (1\to2\to3) + (1\to 3 \to 2)\,.
\label{lead3}
\end{eqnarray}
Performing the derivatives, and 
using the following integration formulas 
\begin{eqnarray}
\label{masterinteg}
 \int \! d^4x \;
\frac{x^{2(N-j)}}{(x^2 + \rho_1^2)^{N+5}} 
= \frac{ \pi^2 (j+2)!(N-j+1)!}{\rho_1^{2j+6}(N+4)!} \, ,
\end{eqnarray}
we find that the right hand side of Eq.~(\ref{lead3}) vanishes.
Therefore, the leading term of the order $1/(X_{12}^2 X_{13}^2)$
vanishes. This means that the expansion starts from the next-to-leading
order, 
\begin{eqnarray}
 H_{\rm pot}^{U(1)} \biggm|_{\rm 3-body}
= 
\frac{-N_c}{128\pi^4 a} 
\; {\cal O}\!
\left(
\frac{(\rho)^4}{X_{12}^2 X_{13}^4}, 
\frac{(\rho)^4}{X_{12}^3 X_{13}^3}, \cdots 
\right).
\end{eqnarray}
where $\cdots$ represents terms obtained by permutation for the indices
$1,2,3$.  
Here the dependence on $\rho_i$ ($i=1,2,3$)
is fixed to be $(\rho)^4$ by a dimensional analysis.
The expectation value of this $(\rho)^4$ at the leading order in large
$N_c$ is given by the classical value given before.
Then, rescaling the coordinates 
back as $X_{12}\rightarrow \lambda^{1/2}X_{12}$ and write it 
as the 3-dimensional inter-nucleon distance $r_{12}$ since we substitute 
the classical value $X^4_i=0$, we obtain, at the leading
order in $1/N_c$,
\begin{eqnarray}
  H_{\rm pot}^{U(1)} \biggm|_{\rm 3-body}
=
\frac{N_c}{\lambda^3}
\; {\cal O}\!
\left(
\frac{1}{r_{12}^2 r_{13}^4}, \frac{1}{r_{12}^3 r_{13}^3}, \cdots 
\right) \,,
\end{eqnarray}
again $\cdots$ represents term obtained by permutation for the indices
$1,2,3$.  

Note that we are working in a regime 
$\lambda^{-1/2} \ll X_{12,13} \ll 1$ in the unit $M_{\rm KK}=1$.
The natural scale for the 2-body force \cite{Hashimoto:2009ys}
is ${\cal O}(N_c/\lambda X_{12}^2)$. So, if we consider a 
natural separation of the nucleons as $X_{ij}\sim 1/M_{\rm KK}$, 
the 3-body force is suppressed compared to the 2-body force.
We conclude that the 3-body force at short range is
small, for baryons carrying classical and equal spin/isospins.


\vspace{2mm}
\noindent
--- {\bf N-body Nuclear Force.} 
We can easily extend the analysis in the previous section to $N$ 'tHooft
instantons. The result for the leading term vanishes again, as we 
explain briefly below.

The quantity necessary for computing $\hat{A}_0$ is
\begin{eqnarray}
 &&\frac{ (\p_M f)^2}{f^2} 
=  
 \left[\frac{4\rho_1^4}{x^6} + \sum_{i=2}^N \frac{4 \rho_1^2 \rho_i^2  
x\cdot X_{1i} }{x^4 X_{1i}^4}
  + \sum_{i=2}^N \frac{4  \rho_i^4}{X_{1i}^6} \right. \nonumber \\
&& \left.   
+ \sum_{i\neq j}\frac{4\rho_i^2 \rho_j^2 X_{1i}\cdot X_{1j}}{X_{1i}^4
X_{1j}^4} 
\right] \left[
1+ \frac{\rho_1^2}{x^2} + \sum_{i=2}^N \frac{\rho_i^2}{X_{1i}^2}
\right]^{-2}\!\!\! . \,
\label{ffex2}
\end{eqnarray}
The expansion analogous to 3-body case implies that 
the leading order of the 
short-range nuclear force 
in proper to $N$-body would be
\begin{eqnarray}
 H_{\rm pot}^{U(1)} \biggm|_{\rm N-body}
= 
\frac{N_c}{\lambda^{N-1}}
\; {\cal O}\!
\left(\prod_{i=2}^N \frac{1}{X_{1i}^2}, \cdots
\right).
\label{N-order}
\end{eqnarray}
Again, we have rescaled back the coordinates to the original
coordinates and $\cdots$ represents permutation terms.
In the previous section, we showed that for $N=3$ this leading
contribution vanishes, for the 'tHooft instantons. In this section, we
prove that for any $N$ this leading contribution vanishes.

First, in the integral (\ref{a0a0}), it is straightforward to obtain
\begin{eqnarray}
&& 
\frac{(\partial_M f)^2}{f^2}
\Box
\frac{(\partial_P f)^2}{f^2}
=
(N-1)! \;
64 \rho_1^8 (-1)^{N-1} \nonumber \\
&& \;\;\;\times\
x^{2(N-4)} (x^2 + \rho_1^2)^{-N-5} \prod_{i=2}^N \frac{1}{X_{1i}^2} 
\sum_{l=1}^N
l (N-l+1)  \nonumber \\
&& \;\;\;\times\left[
(N\!-\!l)(N\!-\!l\!-\!1)\rho_1^4\! -\! 6(N\!-\!l)x^2 \rho_1^2 + 6x^4
\right] \,. 
\label{ffbff2}
\end{eqnarray}
Using the formula (\ref{masterinteg}), 
this can be easily integrated with $\int d^4x$ to give $0$.

In the hamiltonian (\ref{N-order}), there are additional terms 
\begin{eqnarray}
- \int \! d^4x \left(
\frac{4}{x^2} + \sum_{i=2}^N\frac{4}{X_{1i}^2} 
\right)
\Box 
\frac{(\partial_M f)^2}{f^2}.
\end{eqnarray}
coming
from the first term in Eq.~(\ref{last}). 
In the same manner this is shown to vanish at the order 
(\ref{N-order}).
Therefore, we conclude that the leading order $N$-body nuclear force
(\ref{N-order}) vanishes, for arbitrary $N$.
Note that 
the 2-body nuclear force does not vanish at the leading order, 
as $N=2$ computation is exceptional.


\vspace{2mm}
\noindent
--- {\bf Summary and Discussions.} 
Using the SS model of holographic QCD, we have found that the $N$-body
nuclear force at short range ($N\ge3$) is order of 
$N_c(\lambda r^2)^{-N}$,
for nucleons sharing identical {\it classical} spin/isospins. 
This is small compared to the 2-body force 
which is ${\cal O}(N_c(\lambda r^2)^{-1})$ in contrast,
and it leads
to a hierarchy of the $(N+1)$-body / $N$-body ratio 
$V^{(N+1)}/V^{(N)} \sim 1/(\lambda r^2) \ll 1$ for 
$N\geq 3$, in the unit $M_{\rm KK}=1$. 
This suppression is consistent with our empirical knowledge. 

Effects of the
short-range many-body interaction becomes more prominent for
higher-density nuclear matter. Therefore, for physics of neutron stars
and supernovae, for instance, 
properties of $N$-body interactions such as what is
revealed in this letter are important, even if qualitative. 

Our computation is not fully satisfactory since 
the quantum spin/isospin states of each baryon 
have not been incorporated.
The wave function of the classical spin/isospin is a delta-function 
of the $SU(2)$ orientational moduli of the instantons, thus it is
difficult to relate it with the quantum spin/isospins.
Nevertheless, it is quite remarkable that the generic $N$-body nuclear 
force can be obtained by analytic computations. The successful
performance of this computation owes in particular to the simplicity
of the SS model, in contrast to other chiral soliton models.

Furthermore, the theory
on which our computations of the nuclear force is based 
is not a phenomenological model but a theory which has been ``derived''
from large $N_c$ QCD at strong coupling, through the gauge/string
duality. Therefore, in principle, we can try to address theoretically 
what is different from QCD and what is inherited from it.
The present computations 
go beyond the limitations of arguments using universality of 
chiral symmetry breaking. 

Although the model at hand is for large $N_c$ QCD, 
these two properties of
baryons in holographic QCD would be sufficiently strong
motivations for studying holographic QCD and its relation to nuclear
physics further.

\vspace{5pt}
\noindent
{\bf --- Acknowledgment.}
K.H.~is grateful to M.~Oka for helpful discussions.
K.H.~and N.I.~would like to thank KITP, Santa Barbara and 
YITP at Kyoto University.  
N.I.~also thanks Aspen Center for Physics and RIKEN. 
K.H.~and T.N.~are partly supported by
the Grant-in-Aid for Scientific Research (No.~19740125, No.~21340073), of
the Japan Ministry of Education, Culture, Sports, Science and Technology.


\begin{thebibliography}{10}
\bibitem{1} 
S.~C.~Pieper, V.~R.~Pandharipande, R.~B.~Wiringa, and J.~Carlson,
   Phys.\ Rev.\ {\bf C64}, 014001 (2001).
\bibitem{2} A.~Akmal, V.~R.~Pandharipande, and D.~G.~Ravenhall,
	Phys.\ Rev.\ {\bf C58},   1804 (1998).
\bibitem{3} J.~Fujita and H.~Miyazawa, Prog.\ Theor.\ Phys.\ {\bf 17}, 360
	(1957); {\bf 17}, 366    (1957).
\bibitem{4} S.~A.~Coon, M.~T.~Pena, and D.~O.~Riska, 
        Phys.\ Rev.\ {\bf C52},	2925 (1995). 
\bibitem{5} S.~Nishizaki, Y.~Yamamoto, and T.~Takatsuka,
	Prog.\ Theor.\ Phys.\ {\bf 105}, 607 (2001).
\bibitem{Ishii}
See for recent lattice QCD analyses, N.~Ishii, S.~Aoki and T.~Hatsuda,
  Phys.\ Rev.\ Lett.\  {\bf 99}, 022001 (2007)
  [arXiv:nucl-th/0611096].
 
\bibitem{Maldacena:1997re}
  J.~M.~Maldacena,
  Adv.\ Theor.\ Math.\ Phys.\  {\bf 2}, 231 (1998)
  [Int.\ J.\ Theor.\ Phys.\  {\bf 38}, 1113 (1999)]
  [arXiv:hep-th/9711200].

\bibitem{Sakai:2004cn}
  T.~Sakai and S.~Sugimoto,
  Prog.\ Theor.\ Phys.\  {\bf 113}, 843 (2005)
  [arXiv:hep-th/0412141].

\bibitem{Sakai:2005yt}
  T.~Sakai and S.~Sugimoto,
  Prog.\ Theor.\ Phys.\  {\bf 114}, 1083 (2005)
  [arXiv:hep-th/0507073].


\bibitem{Hata:2007mb}
  H.~Hata, T.~Sakai, S.~Sugimoto and S.~Yamato,
  Prog.\ Theor.\ Phys.\ {\bf 117},1157 (2007)
  [arXiv:hep-th/0701280].

\bibitem{Hata:2007tn}
  H.~Hata and M.~Murata,
  Prog.\ Theor.\ Phys.\  {\bf 119}, 461 (2008)
  [arXiv:0710.2579 [hep-th]].

\bibitem{Hashimoto:2008zw}
  K.~Hashimoto, T.~Sakai and S.~Sugimoto,
  Prog.\ Theor.\ Phys.\  {\bf 120}, 1093 (2008)
  [arXiv:0806.3122 [hep-th]].

\bibitem{Hong:2007kx}
 D.~K.~Hong, M.~Rho, H.~U.~Yee and P.~Yi,
 Phys.\ Rev.\  D {\bf 76}, 061901 (2007)
 [arXiv:hep-th/0701276].

\bibitem{Hashimoto:2009ys}
  K.~Hashimoto, T.~Sakai and S.~Sugimoto,
  Prog.\ Theor.\ Phys.\  {\bf 122}, 427 (2009)
  [arXiv:0901.4449 [hep-th]].

\bibitem{Belavin:1975fg}
  A.~A.~Belavin, A.~M.~Polyakov, A.~S.~Shvarts and Yu.~S.~Tyupkin,
  Phys.\ Lett.\  B {\bf 59}, 85 (1975).



\bibitem{Atiyah:1978ri}
  M.~F.~Atiyah, N.~J.~Hitchin, V.~G.~Drinfeld and Yu.~I.~Manin,
  Phys.\ Lett.\  A {\bf 65}, 185 (1978).

\bibitem{Corrigan:1983sv}
  E.~Corrigan and P.~Goddard,
  Annals Phys.\  {\bf 154}, 253 (1984).

\bibitem{Osborn:1979bx}
  H.~Osborn,
  Nucl.\ Phys.\  B {\bf 159}, 497 (1979).


\end{thebibliography}
\end{document}